\documentclass[%
 amsmath,amssymb,showkeys,11pt
]{revtex4-1}
\usepackage{graphicx}% Include figure files
\usepackage{dcolumn}% Align table columns on decimal point
\usepackage{bm}% bold math
\usepackage{algorithm}            % For \mathbb
\usepackage{psfrag,epsfig}
\usepackage{slashbox}
\usepackage{color}

\usepackage[usenames,dvipsnames,svgnames,table]{xcolor}

\newcommand{\cben}[1]{#1}

\begin{document}

\preprint{AIP/123-QED}

\title[Sensitivity analysis and study of the mixing uniformity of a microfluidic mixer]{Sensitivity analysis and study of the mixing uniformity of a microfluidic mixer}% Force line breaks with \\

\author{Benjamin Ivorra}
\email{ivorra@mat.ucm.es (Communicating Author)}
 \affiliation{Dept. de Matem\'{a}tica Aplicada \& Instituto de Matemática Interdisciplinar, Universidad Complutense de Madrid, Plaza de Ciencias, 3, 28040 Madrid, Spain.}%Lines break automatically or can be forced with \\

\author{Juana L\'{o}pez Redondo}%
\affiliation{Dept. de Inform\'atica \& Campus de Excelencia Internacional Agroalimentario (ceiA3), Universidad de Almer\'{\i}a.  Ctra. Sacramento, s/n, 04120, Almer\'{\i}a, Spain.}%

%\author{Pilar M. Ortigosa}
%\affiliation{Dept. de Inform\'{a}tica, Universidad de Almeria, ceiA3, Ctra. Sacramento, La Ca\~{n}ada de San Urbano,  04120 Almer\'{i}a, Spain.}%
\author{Angel M. Ramos}
 \affiliation{Dept. de Matem\'{a}tica Aplicada \& Instituto de Matemática Interdisciplinar, 
Universidad Complutense de Madrid, Plaza de Ciencias, 3, 28040 Madrid, Spain.}
\author{Juan G. Santiago}
\affiliation{Mechanical Engineering Dept., Stanford University, 440 Escondido Mall, Stanford, CA. 94305-3030, U.S.A.%\\~
%\\ {\upshape \sffamily Communicating Author: Benjamin Ivorra}
}%Lines break automatically or can be forced with \\
%\email{ivorra@mat.ucm.es}

\date{\today}% It is always \today, today,
\begin{abstract}
We consider a microfluidic mixer based on hydrodynamic focusing, which is used to initiate the folding process of individual proteins. The folding process is initiated by quickly diluting a local denaturant concentration, and we define mixing time as the time advecting proteins experience a specified to achieve a local 
drop in denaturant concentration.  In previous work, we presented a minimization of mixing time which considered optimal geometry and flow conditions, and achieved a design with a predicted mixing time of 0.10 $\mu$s. The aim of the current paper is twofold. First, we explore the sensitivity of mixing time to key geometric and flow parameters. In particular, we study the angle between inlets, the shape of the channel intersections, channel widths, mixer depth, mixer symmetry, inlet velocities, working fluid physical properties, and denaturant concentration thresholds. Second, we analyze the uniformity of mixing times as a function of inlet flow streamlines.  We find the shape of the intersection, channel width, inlet velocity ratio, and asymmetries have strong effects on mixing time; while inlet angles, mixer depth, fluid properties, and concentration thresholds have weaker effects. Also, the uniformity of the mixing time is preserved for most of the inlet flow and distances of down to within about 0.4 $\mu$m of the mixer wall.  We offer these analyses of sensitivities to imperfections in mixer geometry and flow conditions as a guide to experimental efforts which aim to fabricate and use these types of mixers. Our study also highlights key issues and provides a guide to the optimization and practical design of other microfluidic devices dependent on both geometry and flow conditions.

\end{abstract}

%Use showkeys class option if keyword
\maketitle

\section{Introduction \label{intro}}

Protein folding studies \cite{Dunbar} pose a great challenge to microfluidic mixers. Proteins are composed of chains of amino acids which take on complex three-dimensional (3D) structures to achieve a wide range of biological functions \cite{Berg,Roder}. The range of applications of protein folding in research and industry is wide  and includes drug discovery, DNA sequencing, and molecular analysis or food engineering \cite{Gaudet,inf,Russell}. One of the most versatile methods of initiating the process of protein folding is using changes in chemical potential (e.g., changing the concentration of a chemical species) \cite{Park, Yamaguchi, Mansur}.

In this work, we consider a class of microfluidic mixers based on diffusion from (or to) a hydrodynamically focused stream.  This type of mixer was initially proposed by Brody et al. ~\onlinecite{Brody}. A geometrical representation of such a mixer is shown in Figure~\ref{curgeoa} (here our optimized mixer of Ref. \onlinecite{POF1}). The basic features of the design are as follows:  It is composed of three inlet channels and a common outlet channel, and the geometry has a symmetry with the center channel. Typically, a mixture of unfolded proteins and a chemical denaturant solution is injected through the center channel and exposed to background buffers (no denaturant)  streams through the two side channels.  The design goal is to rapidly decrease the denaturant concentration in order to rapidly initiate protein folding in the outlet channel \cite{Dunbar}. Since the publication of Brody et al., there have been significant advances on the design of these mixers \cite{hert,refmixer,yao07} including reduction in consumption rate of reactants, methods of detection, manufacturing and, perhaps most importantly, drastic reductions of the mixing time (i.e. the time required to reach a sufficiently low denaturant concentration).

%For example, while the original mixer of Brody et al. \cite{Brody} showed mixing times greater than 10 $\mu$s (given the mixing measures used here), Hertzog et al. \cite{hert} obtained mixing times of 1.2 $\mu$s. Furthermore, Hertzog et al. \cite{hert,refmixer} and Yao and Bakajin \onlinecite{yao07} pointed out the importance of 3D flow effects and flow inertia in the designs of these mixers but, due to computational limitations, they considered only 2D flow models.

%We consider the microfluidic hydrodynamic focusing mixer initially proposed by Brody et al. in Ref.~\onlinecite{Brody}. As shown in Figure \ref{curgeoa}, this kind of mixer is composed of three inlet channels and a common outlet channel. It is symmetric with respect to its center channel. In the center inlet channel, a mixture of unfolded proteins and a chemical denaturant is injected, whereas in the two side inlet channels, a background buffer is introduced. The objective is to rapidly decrease the denaturant concentration in order to initiate protein folding in the outlet channel \cite{Dunbar}.

%Focusing on this last aspect, the lower the mixing time, the higher the proportion of folded proteins in the outlet stream \cite{Dunbar}.

\begin{figure}
\centering
\psfig{file=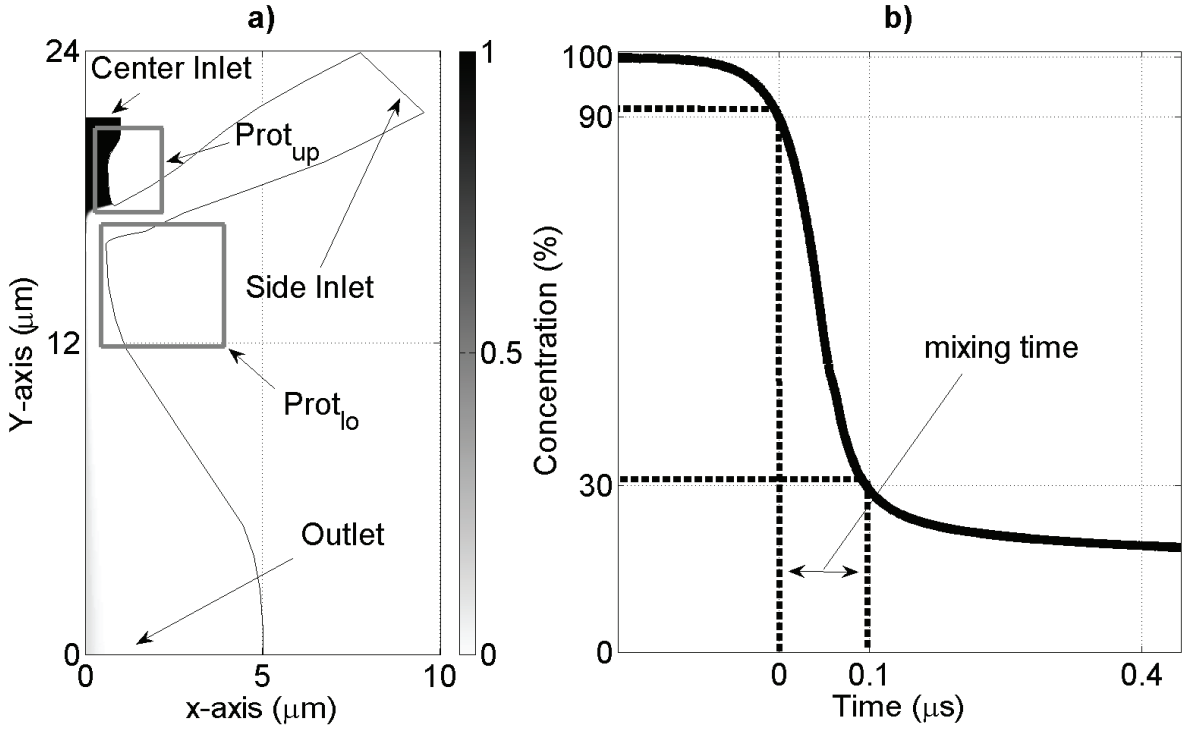,width=16cm} 
\caption{Optimized mixer of Reference \onlinecite{POF1} and associated mixing performance: (a) top view representation of the half of the mixer shape (symmetry with respect to $x$=0 $\mu$m ) of the mixer with a superposed grey scale plot of the denaturant normalized concentration distribution $c$ at width $z$=0 $\mu$m and (b) the time evolution of the denaturant concentration of a particle flowing along  the symmetry
streamline starting from  $x$=0 $\mu$m,  $y$=5 $\mu$m and  $z$=0 $\mu$m. \cben{The parts of the mixer shape corresponding to  the protuberances Prot$_{\rm up}$ and Prot$_{\rm lo}$, introduced in Section \ref{shape}, are highlighted in sub-figure (a)}.\label{res2D}} 
\end{figure}

We recently studied the optimization of the shape and flow conditions of a particular hydrodynamic focused microfluidic mixer\cite{POF1}. The objective was to improve the mixing time of the best mixer designs found in literature, which exhibited mixing times of approximately 1.0 $\mu$s. To this end, we introduced a mathematical model which computes the mixing time for a given mixer geometry and injection velocities. Then, we defined the corresponding optimization problem and solved it by considering a hybrid global optimization method \cite{ijnme,jota,jogo}. This approach was carried out and presented using both 2D and 3D models. To save on computational time, much of the optimization process was conducted using the 2D model. However, our earlier work also pointed out that certain important effects 
 (including the impact of upper and lower mixer walls and inertial effects on the velocity field) can be appreciated only with the 3D model. We therefore performed 3D model studies to analyze such effects. The optimized mixer generated by our approach achieved a mixing time of about 0.10 $\mu$s. The shape of this optimized microfluidic mixer, its concentration distribution and the concentration evolution of a particle in its central streamline are summarized in Figure ~\ref{res2D}. The optimized side and center channel injection velocities were $u_s=$5.2 m s$^{-1}$ and $u_c=$0.2 m s$^{-1}$, respectively.
The optimization problem studied in this previous work was identified as highly nonlinear\cite{redondo1}. Further, the process has many parameters which are difficult to know with great precision in experiments. Therefore, it is important to understand and quantify the stability of the device performance  with respect to these parameters.  This enables identification of key parameters and so guide experimental efforts.  To this aim, we have analyzed the optimized mixer to study and quantify its robustness to parameters perturbations.

We previously presented a very simple sensitivity analysis~\cite{POF1}. That preliminary sensitivity study consisted of random perturbations of all the parameters by taking uniform variations within  a range of \textcolor{black}{$[-\beta \%,+\beta \%]$} of their value. Results showed that the mixing time variations were of the same order as the normalized perturbations considered, suggesting the optimized solution was fairly stable. See Ivorra et al~\cite{POF1} for further information.

Here, we \cben{significantly increase }the scope of our sensitivity analysis.  We quantify the impact of mixing time on the key design parameters of the mixer. The objective of our study is to provide recommendations and guidelines for the fabrication of the device introduced here. More precisely, we consider and study (i) Geometrical parameters defining the mixer shape: the angle defined by the channel intersection, the shape of the channel intersection, the width of the inlet and outlet channels, the mixer depth and possible irregularities in the symmetry of the shape;(ii) Central and side injections velocities; and (iii) Physical coefficients associated with the working fluid and the concentration thresholds of the mixing time definition.

In addition to those sensitivity analysis experiments, we also analyze the uniformity of the mixing time as a function of the inlet streamline location in the inlet channel. This mixing time uniformity analysis quantifies the robustness of the mixing time through the whole inlet flow, and helps place a statistical confidence on observed mixing times. In particular, it helps quantify the so-called wall effect (due to the no-slip condition at the mixer walls, resulting in low velocity values near the walls) on mixer performance.   

The last two decades have seen a large number of microfluidic device designs and their use in a wide range of applications.  Most, if not all, of these devices have performance specifications which are dependent on their geometry and flow control conditions (e.g., flow rates, pressure, inlet concentrations).  Despite this, the systematic study of how performance depends on intentional or untintential design parameters is rarely if ever demonstrated.  For this reason, we also offer the current work as a case study describing the significant challenge and complexity of determining design robustness for microfluidics.

This article is organized as follows: Section \ref{model} introduces the 3D model used to estimate the mixing times. Section \ref{uniform} describes the mixing time uniformity analysis and the results. Section \ref{sensa} presents the numerical experiments carried out to perform the extended sensitivity analysis and deduce major conclusions and design guidelines.

\section{Microfluidic mixer modeling \label{model}} %MODIFY BY JUANI

\begin{figure}[!ht]
\centering
\psfig{file=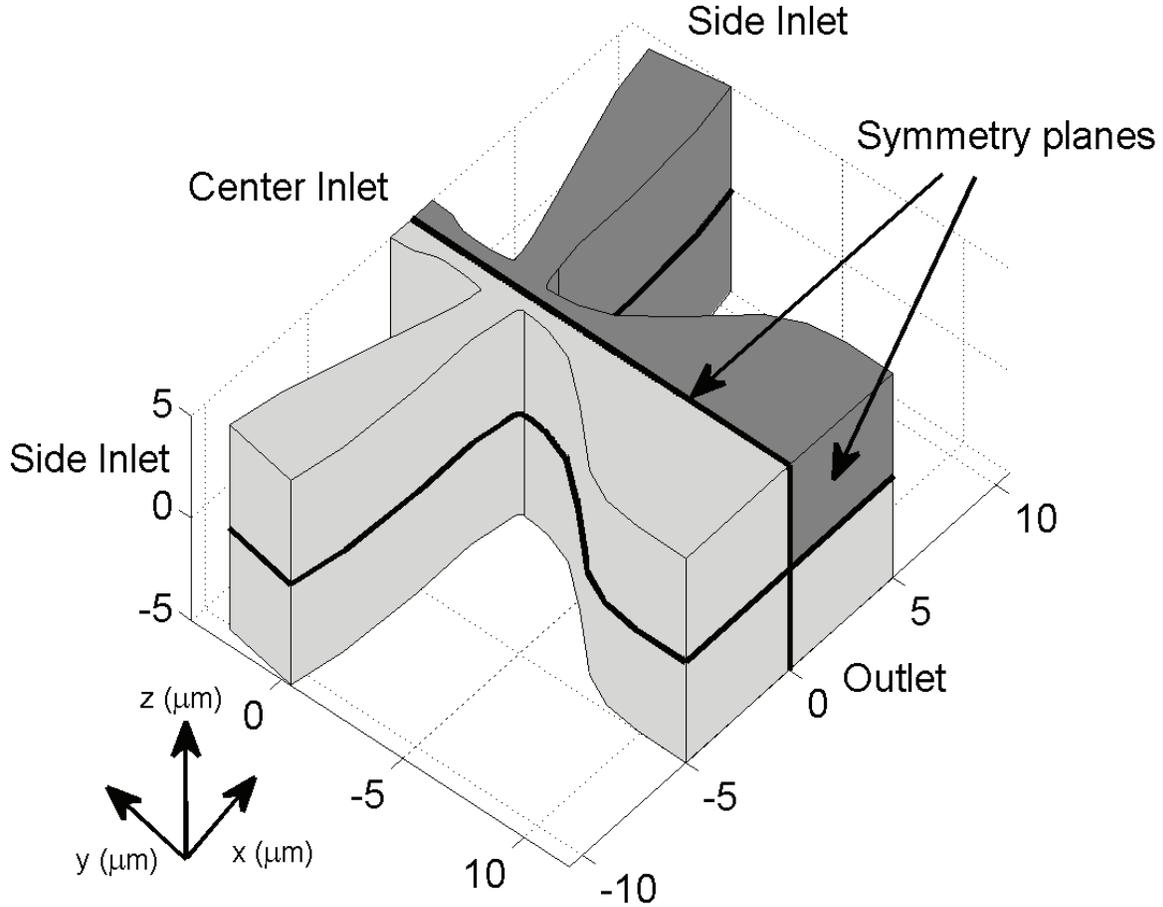,width=16cm} 
\caption{Typical three-dimensional representation of the microfluidic mixer geometry. \cben{The mixer design hydrodynamically focuses a center inlet stream using two side inlets.}  In dark gray we represent the domain $\Omega$ used for numerical simulations. The geometry's \cben{two} symmetry planes are also highlighted. \label{curgeoa}} 
%\textbf{b)} Top view of the domain $\Omega_{s}$ and parameterization of the microfluidic mixer considered for the optimization process.}
\end{figure}

We consider the microfluidic mixer described in Section \ref{intro}. The geometry has two symmetry planes which we use to reduce the simulation domain to a quarter of the mixer. This reduced domain is denoted by $\Omega$, as depicted in Figure \ref{curgeoa}.  The mixer shape is composed of interpolated surfaces, and the inlet velocities are described by a set of parameters denoted by $\phi$, detailed in Ref. \onlinecite{POF1}.  

We consider guanidine hydrochloride (GdCl) as the denaturant \cite{Dunbar,Kawahara}. We assume the mixer liquid flow is incompressible \cite{refmixer}. Thus, the flow velocity and the denaturant concentration distribution are approximated by using the steady configurations of the incompressible Navier-Stokes equations coupled with the convective diffusion equation. More precisely, we consider the following system \cite{massey,hert,refmixer}:

\begin{equation}
\left \{
\begin{array}{ll}
-\nabla \cdot ( \eta ( \nabla  \mathbf{u} +( \nabla  \mathbf{u})^{\top} ) -p\mathbf{I}) + \rho (  \mathbf{u}
\cdot \nabla)  \mathbf{u} = 0& {\rm in} ~~\Omega,\\
\nabla \cdot  \mathbf{u} =0 & {\rm in} ~~\Omega,\\
\nabla \cdot ( -D \nabla c) +  \mathbf{u} \cdot \nabla c = 0& {\rm in} ~~ \Omega,
\end{array}
\right. \label{eql1}
\end{equation}
where $c$ is the denaturant normalized concentration distribution, $\mathbf{u}$ is the flow velocity vector (m s$^{-1}$),  $p$ is the pressure field (Pa), $D=2\times 10^{-9}$ is the diffusion coefficient of the denaturant solution in the background buffer (m$^{2}$ s$^{-1}$), $\eta=9.8 \times 10^{-4}$ is the denaturant solution dynamic viscosity (kg m$^{-1}$ s$^{-1}$) and $\rho=1010$  is the denaturant solution density (kg m$^{-3}$ ).

System (\ref{eql1}) is completed by the following boundary conditions:

For the flow velocity $\mathbf{u}$:
\begin{equation}
\left \{
\begin{array}{ll}
\mathbf{u}=0 & \hbox{on } \Gamma_{w},\\
\mathbf{u}=- u_s {\rm para_1} \mathbf{n} & \hbox{on } \Gamma_{s},\\
\mathbf{u}=- u_c {\rm para_2} \mathbf{n} &\hbox{on } \Gamma_{c},\\
p=0 \hbox{ and }   ( \eta ( \nabla  \mathbf{u} +( \nabla  \mathbf{u})^{\top} )) \mathbf{n}=0 &\hbox{on } \Gamma_{e},\\
\mathbf{n} \cdot \mathbf{u}=0 \hbox{ and } \mathbf{t} \cdot ( \eta ( \nabla  \mathbf{u} +( \nabla  \mathbf{u})^{\top} ) -p\mathbf{I}) \mathbf{n}=0 &\hbox{on } \Gamma_{a},\\
\end{array}
\right. \label{boundcu}
\end{equation}
where $\Gamma_{c}$, $\Gamma_{s}$, $\Gamma_{e}$, $\Gamma_{w}$ and $\Gamma_{a}$ denote the boundaries representing the central inlet, the side inlet, the outlet, the mixer walls and the symmetry plane, respectively; $u_s$ and $u_c$ are the maximum side and center channel injection velocities (m s$^{-1}$), respectively; ${\rm para_1}$ and ${\rm para_2}$ are the laminar flow profiles, which are equal to 0 in the inlet border and to 1 in the inlet center, of the side and central inlets, respectively \cite{massey}; and $(\mathbf{t},\mathbf{n})$ is the local orthonormal reference frame along the boundary.

For the concentration $c$:
\begin{eqnarray}
\left \{
\begin{array}{ll}
 \mathbf{n} \cdot (-D \nabla c +  c \mathbf{u})= - c_0 \mathbf{u} & \hbox{on } \Gamma_{c},\\
 c =0 &\hbox{on } \Gamma_{s},\\
\mathbf{n} \cdot (-D \nabla c )=0 &\hbox{on } \Gamma_{e},\\
 \mathbf{n} \cdot (-D \nabla c + c \mathbf{u})=0 &\hbox{on } \Gamma_{w}  \cup \Gamma_{a},
 \end{array}
\right. \label{boundcc}
\end{eqnarray}
where $c_0 =1$ is the initial denaturant normalized concentration in the center inlet.

In this work, the mixing time of a particular mixer $\phi$, denoted by $J(\phi)$ is defined as the time required to change the denaturant normalized concentration of a typical Lagrangian stream fluid particle situated in the symmetry streamline at depth $z=0$ $\mu$m from $\alpha \%$ to $\omega \%$. It is computed by: 

\begin{equation}
J(\phi)= \int_{c^{\phi}_{\omega}}^{c^{\phi}_{\alpha}} \dfrac{{\rm  d}y}{\mathbf{u}^{\phi}(y)},% \cdot \mathbf{t}},
\label{mtcf}
\end{equation}
where  $\mathbf{u}^{\phi}$ and $c^{\phi}$ denote the solution of System (\ref{eql1})-\eqref{boundcc}, when considering the mixer defined by $\phi$; and  $c^{\phi}_{\alpha}$ and $c^{\phi}_{\omega}$ denote the y-coordinate of points situated along the streamline defined by the intersection of the two symmetry planes $z=0$ $\mu$m and $x=0$ $\mu$m, i.e. the y-axis, where the denaturant normalized concentration $c^{\phi}$ is  $\alpha$ and $\omega$, respectively. By default, we assume  $\alpha=90$\% and $\omega=30 \%$.

The numerical model used to approximate the solutions of System (\ref{eql1})-(\ref{boundcc}) and to compute (\ref{mtcf}) was implemented by coupling Matlab scripts with COMSOL Multiphysics 3.5a models.

%--------------------------------------------------------------------------
%--------------------------------------------------------------------------
%--------------------------------------------------------------------------
\section{Uniformity of the mixing time \label{uniform}}

We first analyze the non-uniformity of mixing times across the focused stream for our optimized mixer $\phi_o$. Indeed, as suggested in Refs. \onlinecite{Park2,yao07}, the mixing time can be measured not only in the symmetry streamline, situated on the (x,z)=(0,0) segment, but also in other streamlines. We are interested in the uniformity of mixing times, as protein states in these mixers are quantified \cben{experimentally} within a finite probe volume which integrates signal \cben{throughout} a volume in space within the mixing region. \cben{This measurement volume is} fed in principle by all streamlines of the center inlet channel. 

We consider 100 streamlines, denoted by $(sl_{i,j})_{i,j=1}^{10}$, starting from a finite set of points, which are denoted by $\Sigma_{\Gamma_{c}}$, in $\Gamma_{c}$. Here, $\Sigma_{\Gamma_{c}}= \{ P_{(i,j)} | i=1,...,10 $ and $ j=1,...,10 \}$ where $P_{(i,j)}= (\frac{i}{10} 0.9\mu$m$, \frac{j}{8} 0.75 \mu$m$)$. In the previous definition, the maximum coordinate in the x-axis (i.e., 0.9 $\mu$m) has been selected in order to avoid particles too close to the wall $\Gamma_{w}$,  and the maximum coordinate in the z-axis (i.e., 0.75 $\mu$m) has been chosen as a characteristic 1.5 $\mu$m  depth of field for confocal microscope imaging (i.e., extent of the measurement volume)\cite{hert}. Those streamlines are numerically approximated by considering an explicit Euler scheme and the velocity vector $\mathbf{u}$ obtained by solving System \eqref{eql1}-\eqref{boundcc} \cite{inf}.

For each streamline $sl_{i,j}$, we compute the associated mixing times, denoted by $t_{sl_{i,j}}$, in a manner similar to Equation \eqref{mtcf}. More precisely, $t_{sl_{i,j}}$ is defined as the time required by a protein within a Lagrangian fluid particle to travel from $c^{sl_{i,j}}_{90}$ to $c^{sl_{i,j}}_{30}$, where $c^{sl_{i,j}}_{90}$ and $c^{sl_{i,j}}_{30}$ denote the points within $sl_{i,j}$ with a concentration of $90$\% and $30$\%, respectively. Next, we study the spatial distribution according to the streamline starting point in $\Sigma_{\Gamma_{c}}$, the maximum value, the mean value and the standard deviation of  $(t_{sl_{i,j}})_{i,j=1}^{10}$. Furthermore, we also compute the weighted mixing time value of $sl_{i,j}$, denoted by $\overline{t_{sl_{i,j}}}$ and defined as
\begin{equation} \label{mmt}
\overline{t_{sl_{i,j}}}=\frac{\omega_{i,j} t_{sl_{i,j}}}{\sum_{i,j=1}^{10} \omega_{i,j}},
\end{equation}
where $\omega_{i,j}$ denotes the velocity of a particle in the streamline $sl_{i,j}$ at its initial position $x^{\rm init}_{(i,j)}$. This choice of weight coefficients reflects the fact that the probe volume used to measure experimentally the mixing time receive particles more frequently from streamlines with the highest velocities. The maximum and standard deviation values of those weighted mixing times $(\overline{t_{sl_{i,j}}})_{i,j=1}^{10}$ are also studied.

Furthermore, due to the fact that the depth of the mixer is 10 times larger than the minimum width of the center channel, the mixing time variations in the z-axis direction are negligible in comparison to the variations in the x-axis \cite{hert}. Thus, we perform a more extensive uniformity analysis along the x-axis, by considering 100 streamlines, denoted by $(sl_{i,z=0})_{i=1}^{100}$, in the plane $z=0$ starting from the set of points $P_{(i,j)}= (\frac{i}{100} 0.9\mu$m$, 0 \mu$m$)$ in $\Gamma_{c}$. The methodology is the same as that introduced previously. In this case, we also compute the evolution of both the mean value and standard deviation of $(t_{sl_{i,z=0}})_{i=1}^{k}$ and $(\overline{t_{sl_{i,z=0}}})_{i=1}^{k}$, with $k=1,...,100$. These results will be compared with the ones presented in Ref. \onlinecite{hert}.

Our study of mixing time uniformity yielded that the mean mixing time value obtained by considering $(t_{sl_{i,j}})_{i,j=1} ^{10}$ was 0.34 $\mu$s with a standard deviation of 0.17 $\mu$s. As expected, the maximum mixing time value was reached at the streamline $sl_{10,10}$ with a value of 1.43 $\mu$s.

\begin{figure}
\begin{center}
    \includegraphics [width=11cm]{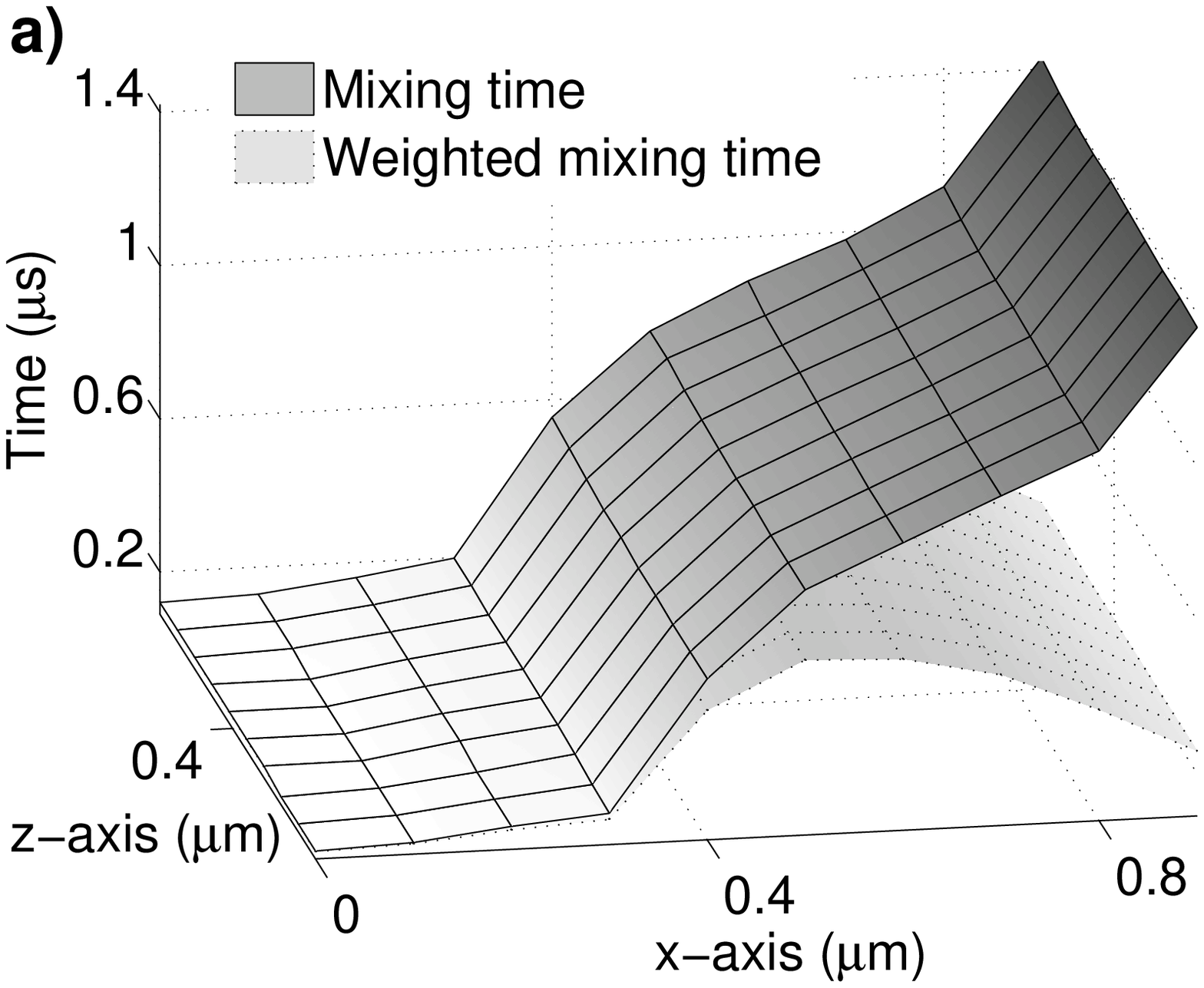}\\
    \includegraphics [width=11cm]{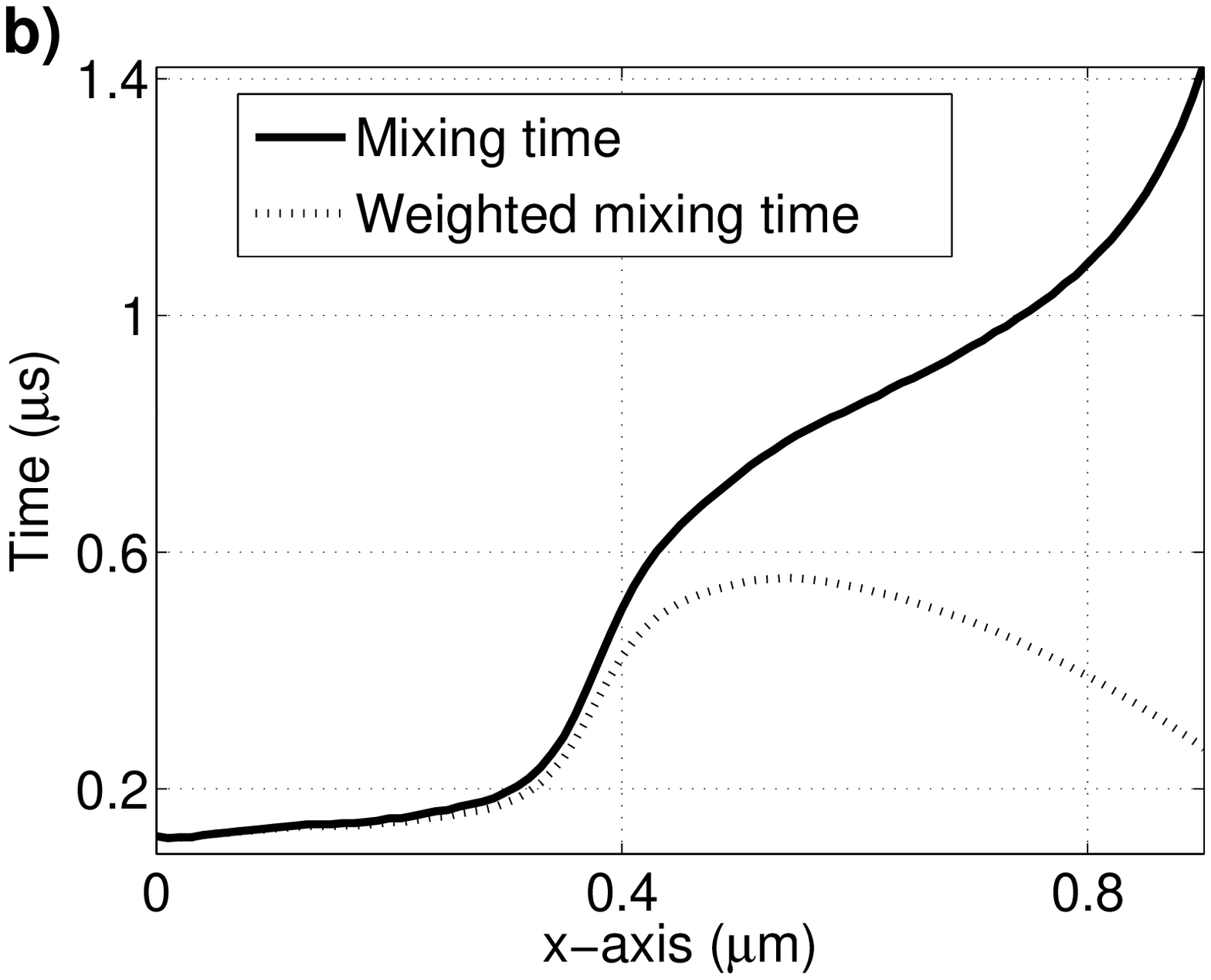}
\end{center}
\caption{Results obtained during the analysis of the mixing time non-uniformity for the optimized mixer: \textbf{a)} Mixing and weighted mixing times obtained according to the position $(x,z)$ in $\Gamma_c$ of the initial particle for the considered streamlines, \textbf{b)} Mixing and weighted mixing times obtained as a function of the position  $x$ in $\Gamma_c$  of the initial particle  and for the streamlines considered in the plane $z=0$. \cben{The weighted mixing time reflects the frequency of events (measurements of proteins along said streamline) as determined by the stream-line averaged velocity.  The lower velocities near the wall yield longer mixing times but are less frequent.}}\label{uan}
\end{figure}

The mixing times $t_{sl_{i,j}}$ and weighted mixing times $\overline{t_{sl_{i,j}}}$  of the considered streamlines $(sl_{i,j})_{i,j=1}^{10}$ are presented in Figure \ref{uan}-(a). As shown, %the mixing time is more sensitive to changes in the X-axis, with a standard deviation of 0.18 $\mu$s in this direction, than in the Z-axis, with a standard deviation of 0.01 $\mu$s. Furthermore, we can see that the velocities of streamlines close to the mixer wall $\Gamma_{w2}$ (i.e., at position $>$0.8 $\mu$m) are low, and their corresponding mixing times are high (i.e., $>$1$\mu$s). However, these near-wall, slow-moving particles contribute only infrequently to probe volume detection events. 
within 0.4 $\mu$m of the centerline, the mixing times vary between 0.1 $\mu$s and 0.5 $\mu$s, and this region accounts for 60\% of the detection events (i.e., considering the sum of the weight coefficients $(\sum_{i=1}^{4}\sum_{j=1}^{10} \omega_{i,j})/(\sum_{i,j=1}^{10} \omega_{i,j})$).  In contrast, the near-wall region of [0.7, 0.9] $\mu$m of the centerline have mixing times between 1 and 1.43 $\mu$s, but these streamlines contribute to only 10\% of detection events (i.e., considering $(\sum_{i=7}^{10}\sum_{j=1}^{10} \omega_{i,j})/(\sum_{i,j=1}^{10} \omega_{i,j})$). 

Next, the mixing times $t_{sl_{i,z=0}}$ and weighted mixing times $\overline{t_{sl_{i,z=0}}}$ across the streamlines $(sl_{i,z=0})_{i=1}^{100}$ are plotted versus spanwise streamline position in Figure \ref{uan}-(b). For these 100 streamlines, the mean mixing time computed by considering $(t_{sl_{i,z=0}})_{i=1}^{100}$ was  0.32 $\mu$s with a standard deviation of  0.16 $\mu$s. Again, we can observe that particles near the walls exhibit higher mixing times ($>$1 $\mu$s). However, these near-wall-slow-moving particles contribute only infrequently to probe volume detection events. 

%The maximum mixing time is 1.41 $\mu$s. The mixing time and weighted mixing time distributions, regarding the initial position of the streamlines, is depicted in Figure \ref{uan}-b.  From the evolution of the mixing time value according to the wall distance, it is possible to observe that the streamlines starting from points at distances of  $\sim$0.3 $\mu$m of the channel center, present similar mixing times, inferior to $0.2$ $\mu$s, with a small standard deviation of 0.01 $\mu$s. For other streamlines starting from points of X-axis $>$0.3 $\mu$m, the mixing times increase considerably as we approach to the wall $\Gamma_{w}$. 

We note similar phenomena were reported  in Ref. \onlinecite{hert}. However, the mixer presented in that work exhibited a mean mixing time, considering streamlines in the plane z=0, of 3.1 $\mu$s with a standard deviation of 1.5 $\mu$s. The maximum mixing time value was 10 $\mu$s, obtained for the streamline closer to the wall $\Gamma_{w}$. The optimized mixer design presented here therefore offers better mixing time uniformity leading to more consistent measurements and less scatter in measurement ensembles.

%--------------------------------------------------------------------------
%--------------------------------------------------------------------------
%--------------------------------------------------------------------------
\section{Sensitivity analysis of the model parameters \label{sensa}}

We here present a study of the influence of key parameters of the model described in Section \ref{model} on mixer mixing time. We vary parameters individually, fixing the values of others to the corresponding value of the optimized mixer $\phi_o$. We note that, in our previous work, we explored the impact of simultaneous perturbations on the whole set of parameters on mixer performance\cite{POF1}. We here perform the more complete influence of individual perturbations on the mixing time.  We believe such individual parameter perturbation analyses are also more useful to designers in identifying key parameters and methods for fabrication.  We consider  the following percent variation function:		
\begin{equation}
E(\phi_{p})=100\dfrac{| J(\phi_o)-J(\phi_{p})|}{J(\phi_o)}.\label{err}
\end{equation}
where $\phi_{p}$ represents the perturbed mixer.

The parameters analyzed can be classified in three categories: (i) geometrical parameters defining the mixer shape; (ii) central and side injections velocities; and (iii) physical coefficients associated with the denaturant solution and the concentration threshold in the mixing time definition.

%{\bf Additionally, a final study has been performed by perturbing all the previous parameters as well as the symmetry plane of the mixer. }

%--------------------------------------------------------------------------
%--------------------------------------------------------------------------
\subsection{Geometrical parameters}
In the following computational experiments, we analyze the variation on the mixing time due to changes in: (i) the angle defined at the channel intersection; 
(ii) the shape of the channel intersection; (iii) the width of the inlet and outlet channels; (iv) the mixer depth; and (v) perturbation in the symmetry of the mixer shape.

%--------------------------------------------------------------------------
\subsubsection{Inlet intersection angle}
\begin{figure}[!ht]
\centering
\psfig{file=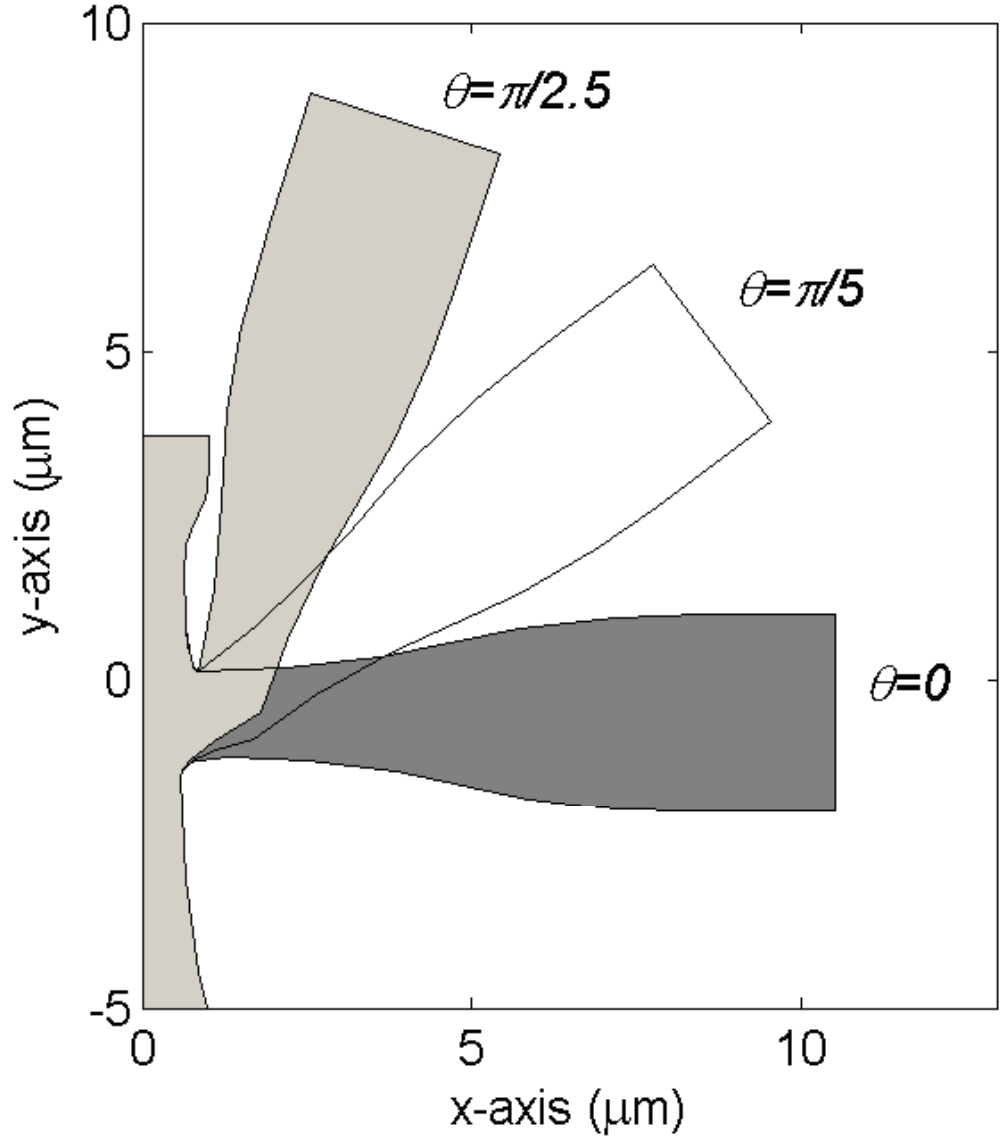,width=16cm} 
\caption{\cben{Three mixer}  shapes \cben{for inlets} intersection angles of $\theta=0$ (light grey), $\theta=\pi/5$ (white) and $\theta=2\pi/5$ (dark grey). Only the area where the shape changes is \cben{shown}. \label{angf}} 
\end{figure}

First, we study the angle between the x-axis and the mixer side channel, denoted by $\theta$. The optimized value $\theta=\pi/5$ is varied from 0 up to $2\pi/5$ by considering 50 equally spaced intermediate values (i.e, we perform 50 evaluations of our model). A geometrical representation of those variations is showed in Figure~\ref{angf}. 

%We denote $\theta$ as the angle between the X-axis and the mixer side channel. In the optimized shape, its value is equal to  $\theta=\pi/5$. We vary this angle from 0 up to $\pi/(2.5)$ by considering 50 equispaced intermediate values. A geometrical representation of those variations is presented in Figure \ref{angf}.

\begin{figure}[!ht]

\centering
\psfig{file=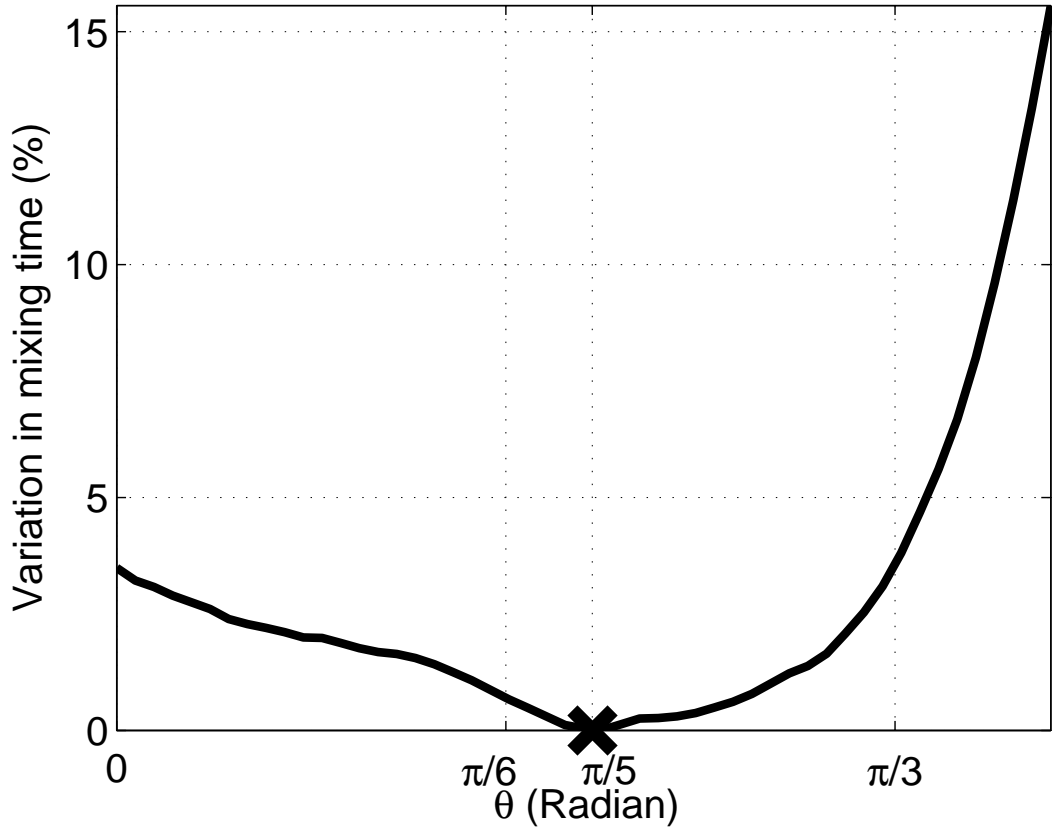,width=16cm} 
\caption{\cben{Percent variation  of mixing time as a function of deviation from the optimal angle $\theta$ value (denoted by X=$\pi$/5, c.f. Figure \ref{angf}).  Mixing time is relatively insensitive to small errors in angle of the side channel.  The distribution shows the strongly non-linear dependence of mixing time on geometry.} \label{angr}} 
%\textbf{b)} Top view of the domain $\Omega_{s}$ and parameterization of the microfluidic mixer considered for the optimization process.}
\end{figure}

Perturbations on $\theta$ have generated a mean variation in the mixing time of 3\%. Figure~\ref{angr} gives a graphical representation of the obtained results. As we can observe on this plot, the maximum variation was around 15\% and was obtained for $\theta=2\pi/5$. Furthermore, the variation was less than 4\% for angles lower than $\pi/3$, and grew up exponentially after that value. This suggests that the angle is not a sensible parameter for the mixer performance. 

%considering angles lower than $\pi/(3)$ generates variation inferior to 4\%, for larger values the variation start to grow up exponentially. However, in all cases those variations can be considered as reasonable as the maximum mixing time stills of the order of 0.1 $\mu$s. This seems to indicate that the angle is not a sensible parameter for mixer performance.

%--------------------------------------------------------------------------
\subsubsection{Shape of the channel intersection \label{shape}}

\begin{figure}[!ht]
\centering
\psfig{file=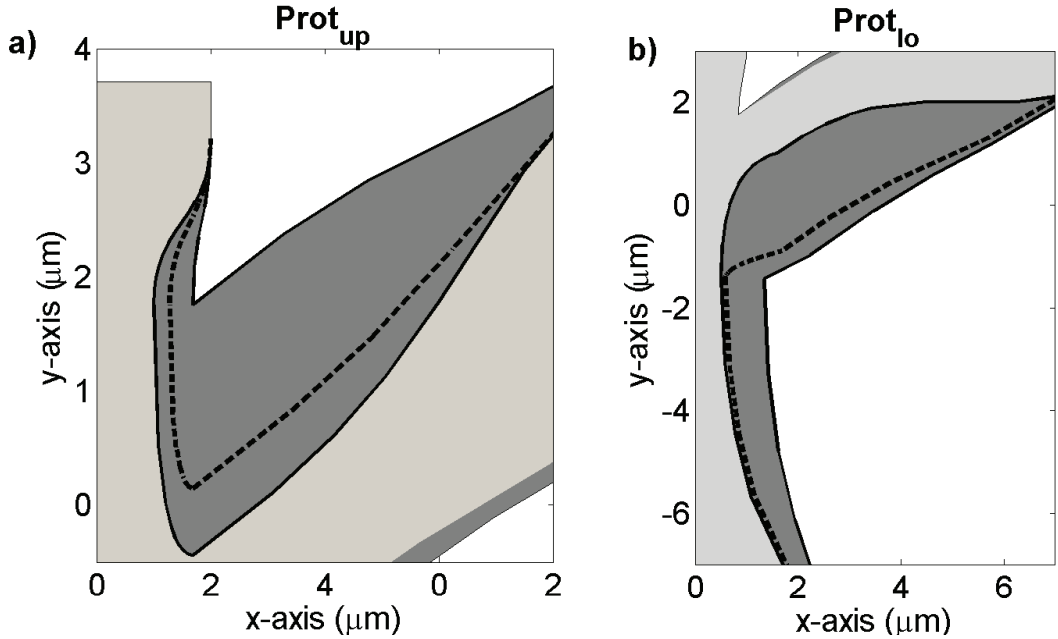,width=16cm} 
\caption{\cben{Mixer shapes highlighting the range of mixer shapes we explored. In both cases, the optimal shape is shown as a dark line.  Shown are detailed views of the shape of the channels intersection near the \textbf{a)} upper (denoted by Prot$_{\rm up}$) and \textbf{b)  lower (denoted by Prot$_{\rm lo}$)} corners of the intersection region.} The dark gray zones correspond to the domain, between the shape of the maximum and the minimum protuberance allowed by the model considered here (according to a minimum channel width of 1$\mu$m). 
\label{cornf}} 
%\textbf{b)} Top view of the domain $\Omega_{s}$ and parameterization of the microfluidic mixer considered for the optimization process.}
\end{figure}

We now study the impact of the shape of the area where the three inlets and the outlet intersect. The shapes allowed by our model are built by considering Beziers curves and describe a 'bubble' (also called protuberance) invading the central and side inlets from the upper corner (according to y-axis) and a protuberance invading the outlet and side inlets from the lower corner. These protuberances are defined according to a restriction (due to a convenient lithographic and plasma etching limitation) of a minimum channel width of 1$\mu$m. For the sake of simplicity, those bubbles are only described by two scalar numbers Prot$_{\rm up}$ and Prot$_{\rm lo}$  in $[0,1]$, where 0 corresponds to the minimum bubble shape and unity is the maximum bubble shape of the upper and lower corner, respectively, as allowed by the model. The optimal shape corresponds to Prot$_{\rm up}$=0.8 and Prot$_{\rm lo}$=0.7. \cben{The parts of the mixer shape corresponding to Prot$_{\rm up}$ and Prot$_{\rm lo}$ are presented in Figure \ref{res2D}.} A geometrical representation of the minimum, maximum and optimal shapes of the protuberances is given in Figure \ref{cornf}. 

\begin{figure}[!ht]
\centering
\psfig{file=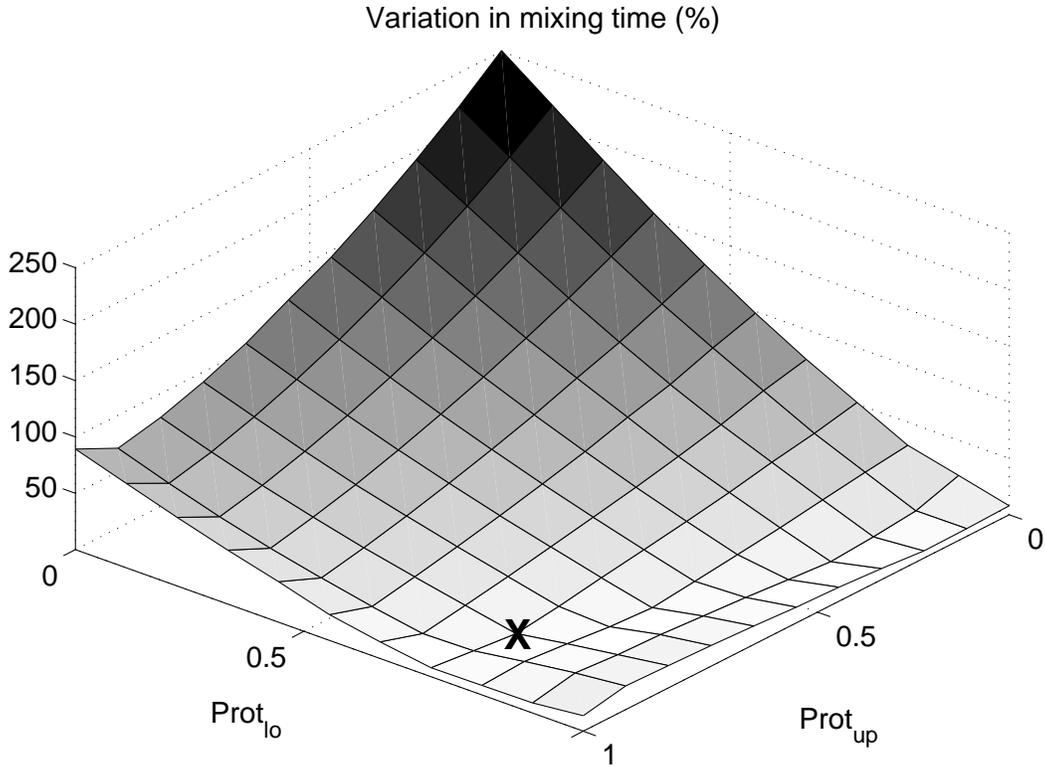,width=16cm} 
\caption{\cben{Percent variation  of} mixing time for the optimal shape (represented by X) for the protuberance magnitudes considered and described in Section \ref{shape}. Protuberance parameter values Prot$_{\rm up}$ and Prot$_{\rm lo}$ each vary from 0 to unity with a grid step size of 0.1. \cben{The details of the protuberance shape of the side channels is an important feature.}    \label{cornr}} 
%\textbf{b)} Top view of the domain $\Omega_{s}$ and parameterization of the microfluidic mixer considered for the optimization process.}
\end{figure}

This experiment consisted of computing the mixing time of the mixer generated by considering all the possible combination of values of Prot$_{\rm up}$ and Prot$_{\rm lo}$ in $[0,1]$ with a grid step size of 0.1. This required 121 evaluations of our model. The variation of the mixing time according to analyzed values of Prot$_{\rm up}$ and Prot$_{\rm lo}$ is presented in Figure \ref{cornr} and values are reported in Table \ref{tcornr}. As shown by both the figure and the table, for values of Prot$_{\rm up}$ and Prot$_{\rm lo}$ lower than 0.5, the mixing time dramatically increased from 50\% up to 250\%. This indicates that a minimum protuberance in both upper and lower corners should be considered in order to obtain an efficient mixing time. Furthermore, when Prot$_{\rm up}$ and Prot$_{\rm lo}$ were greater than 0.5, the variation in mixing time was moderated and was lowered by 22\%, which can be considered as a reasonable value. In addition to those first results, we see that the impact on the mixing of Prot$_{\rm up}$ was greater than Prot$_{\rm lo}$. For instance, by decreasing the parameter Prot$_{\rm lo}$ from 1 to 0 and fixing the value of Prot$_{\rm up}$=1, we have generated mixing time variations up to 50\%, whereas by decreasing the parameter Prot$_{\rm up}$ and fixing Prot$_{\rm lo}$=1 we have obtained a maximum 20\% variation of mixing time. This result is consistent with the fact that the length of the lower corner is much larger than that of the upper (see Figure \ref{cornf}), thus, its influence on the mixing time is expected to be greater.  

\begin{table}
\begin{center}
\caption{Percentage variation  in mixing time of the optimal design value and considering the protuberances described in Section \ref{shape}. Here protuberances parameters  Prot$_{\rm up}$ and Prot$_{\rm lo}$, each vary from 0 to unity with a grid step size of 0.1. The optimal shape (-) is obtained with 
Prot$_{\rm up}$=0.8 and Prot$_{\rm lo}$=0.7. \label{tcornr}}
\vspace{0.5cm}
\begin{tabular}{|c||rrrrrrrrrrr|}
\hline
\backslashbox{\textbf{Prot$_{\rm lo}$}}{\textbf{Prot$_{\rm up}$}}  & 0.0 & 0.1 & 0.2 & 0.3 & 0.4 & 0.5 & 0.6 & 0.7 & 0.8 & 0.9 & 1.0\\
\hline
\hline
0.0 & 251 & 220 & 192 & 167 & 144 & 126 & 106 & 91 	& 79 & 71 & 89\\
0.1 & 218 & 192 & 167 & 145 & 125 & 107 & 90 	& 76 	& 65 & 56 & 75\\
0.2 & 186 & 163 & 142 & 123 & 105 & 89 	& 75 	& 62 	& 51 & 42 & 61\\
0.3 & 155 & 136 & 118 & 102 & 86 	& 73 	& 60 	& 48 	& 39 & 30 & 48\\
0.4 & 125 & 110 & 95 	& 82 	& 69 	& 57 	& 47 	& 37 	& 28 & 20 & 36\\
0.5 & 98 	& 87 	& 74 	& 64 	& 52 	& 43 	& 34 	& 26 	& 18 & 11 & 20\\
0.6 & 74	& 65 	& 54 	& 45 	& 37 	& 29 	& 22 	& 15 	& 9  & 3  & 15\\
0.7 & 51 	& 43 	& 37 	& 29 	& 23 	& 17 	& 11 	& 5 	& -  & 4  & 8\\
0.8 & 30 	& 25 	& 19 	& 14 	& 9 	& 5 	& 1 	& 3 	& 7  & 11 & 9\\
0.9 & 19 	& 7 	& 3 	& 2 	& 3 	& 7 	& 9 	& 12 	& 14 & 17 & 9\\
1.0 & 8 	& 7 	& 6 	& 10 	& 13 	& 14 	& 16 	& 19 	& 20 & 21 & 14\\
\hline
\end{tabular}
\end{center}
\end{table}

\cben{From the previous results, we conclude that the mixing time is sensitive to the shape of these protuberances.}

%--------------------------------------------------------------------------
\subsubsection{Channel width \label{widthe}}

We are here interested in estimating the impact of the inlets and outlet widths on the mixing time (i.e., the minimum width of these channels where the flow they carry first interact with the neighbouring streams). This study is interesting as the mixer design and general shape can be scaled geometrically and inserted into different devices. We note that the channel widths were fixed during the optimization process in Ref. \onlinecite{POF1} and were set to values suited for the mixer implementation and validation studies, as the one carried out in Ref. \onlinecite{hert}.

\begin{figure}[!ht]
\centering
\psfig{file=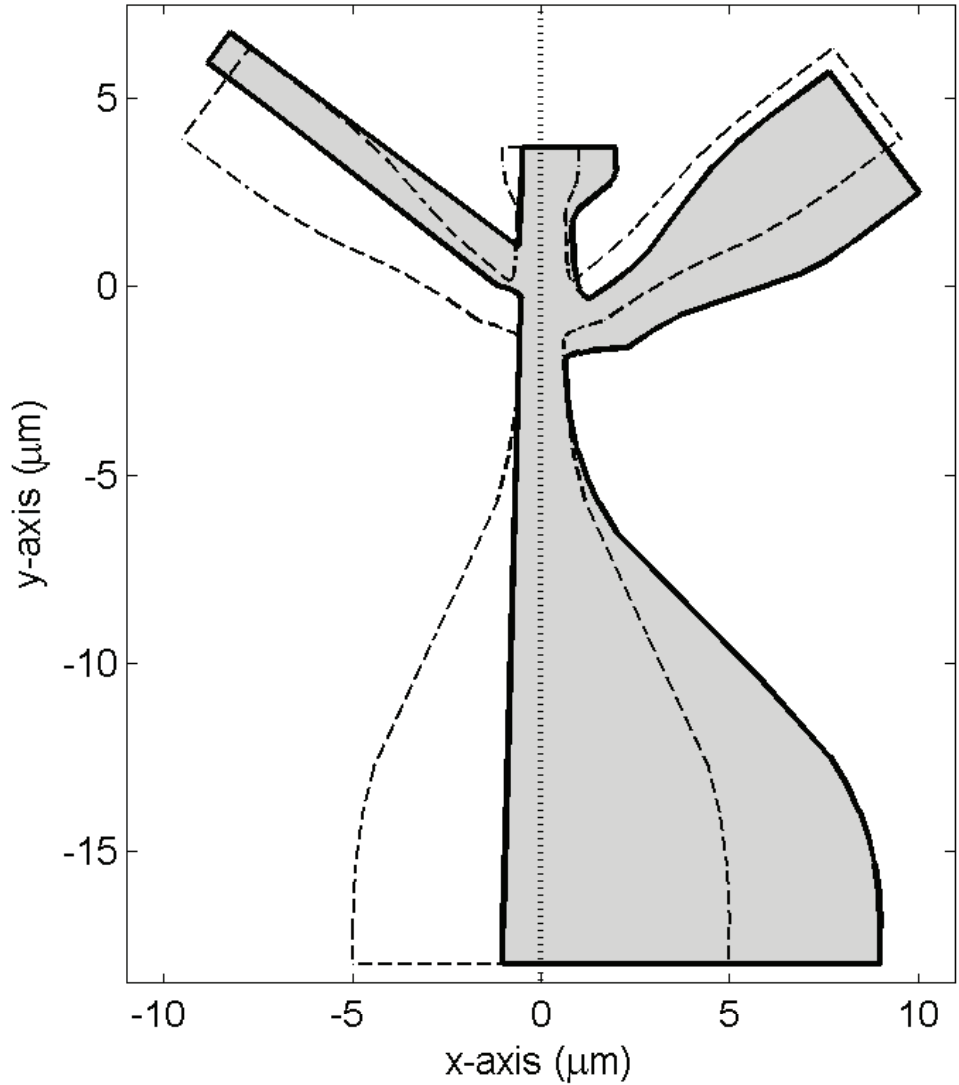,width=16cm} 
\caption{Shape of the mixer for lengths of the channels set to their maximum (continuous line for $x>$0 m) values and minimum values  (continuous line for $x<$0 m) . The dash-dot line corresponds to the axis $x=0$ \cben{(and the y-z symmetry plane)}. The optimal mixer shape is represented by a dashed line. We note \cben{the choice of parameterization of the mixer shape (including side inlet channel width) determines the position of the region corresponding to the channel intersection. Our geometry variations therefore considered a wide range of shapes and relative channel lengths.}
\label{widf}} 
%\textbf{b)} Top view of the domain $\Omega_{s}$ and parameterization of the microfluidic mixer considered for the optimization process.}
\end{figure}

We considered a width denoted by $w_{c}$ $\in$ [1$\mu$m,4$\mu$m]  for the central inlet, a width denoted by $w_{s}$ $\in$ [1$\mu$m,4$\mu$m]  for the side inlets and a width denoted by $w_{o}$ $\in$ [2$\mu$m,18$\mu$m]  for the outlet. The original optimized shape exhibited  $w_{c}=2\mu$m, $w_{s}=3\mu$m  and $w_{o}=10\mu$m. All possible configurations of channel widths were tested by considering a mesh of step size of 1 $\mu$m for each width, which represents a total of 272 evaluations of our model. Representations of the mixer shape with  all channel widths set to their maximum or  minimum values, are depicted by Figure \ref{widf}.

In order to check the importance of each channel width on the mixing time regarding all possible configurations of other width, we considered percent variations denoted by $WE$ and the mean evolution of the mixing time $MET$ according to each width. Both processes are explained below. We illustrate the process of computing $WE$ and $MET$ in the case of $w_{c}$. This approach can be extended to $w_{s}$ and $w_{o}$. 

The value $WE_{w_{c}}(j,k)$ represents a measure of the variation of the mixer mixing time according to changes in $w_{c}$  when other widths are fixed to $w_{s}=j$ and $w_{o}=k$, and is given by
\begin{equation}
WE_{w_{c}}(j,k)=100 \dfrac{1}{4} \sum_{i=1}^{4} \dfrac{|  J(\phi_{i,j,k})- \hbox{mean}_{i} J(\phi_{i,j,k})|}{\hbox{mean}_{i} J(\phi_{i,j,k})}, \label{werr1}
\end{equation}
where $\phi_{i,j,k}$ denotes the mixer obtained by considering $w_{c}=i,w_{s}=j$, $w_{o}=k$ and the other parameters set to the optimal values and $\hbox{mean}_{i} J(\phi_{i,j,k})$ denotes the mean value of the mixing time obtained by varying only $i$. We compute $WE_{w_{c}}(j,k)$ for $j=1,..,4$, $k=2,...,18$ and report its mean, minimum and maximum values according to $j$ and $k$. Those results are reported in Table \ref{widt}.

\begin{figure}[!ht]
\centering
\psfig{file=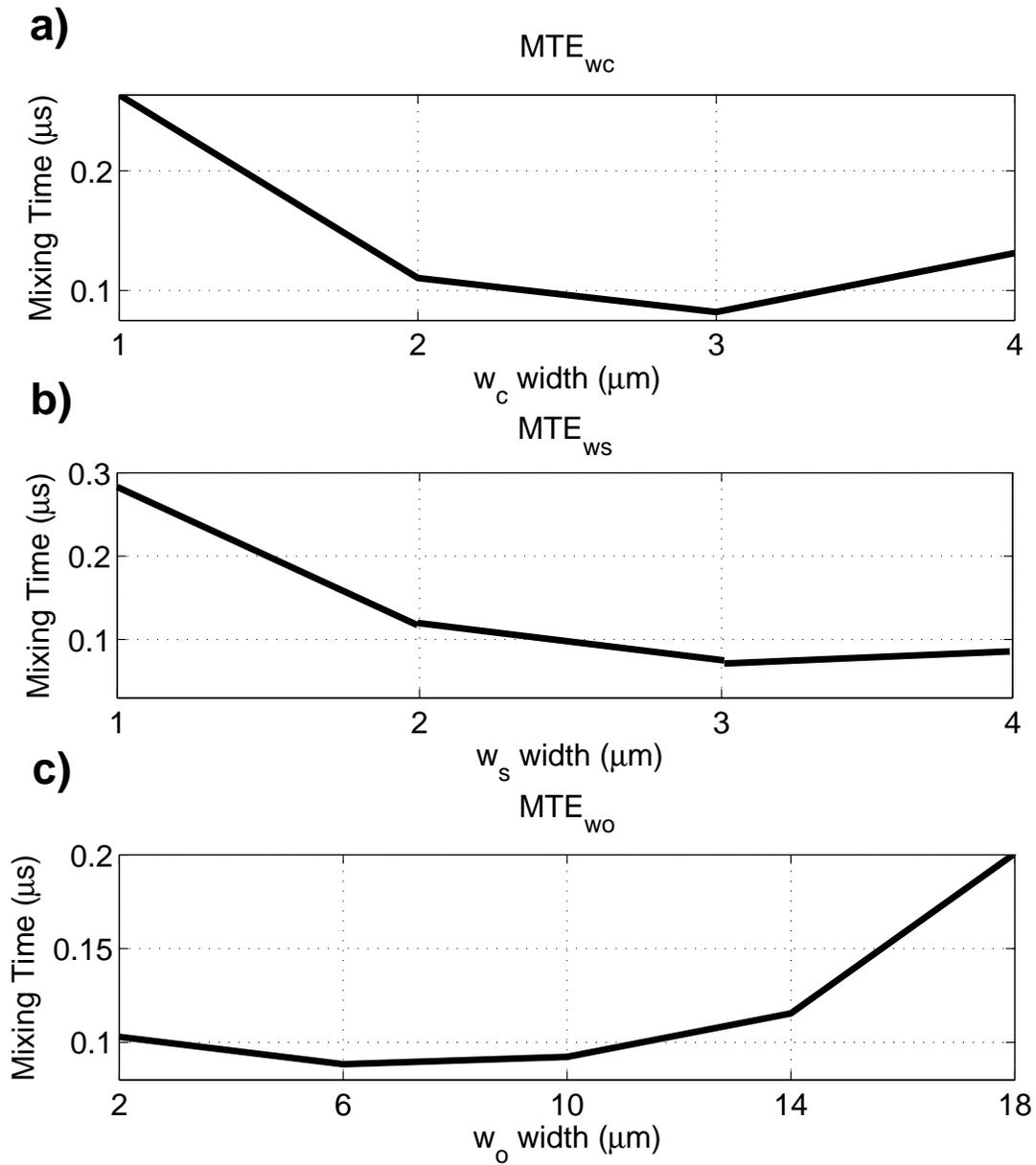,width=16cm} 
\caption{Dependence of mean values of mixing times as defined in Section \ref{widthe} as a function of inlet \cben{and outlet} channel widths. Shown are plots  \textbf{a)} $MTE_{w_{c}}$, \textbf{b)} $MTE_{w_{s}}$ and \textbf{c)} $MTE_{w_{o}}$ %as a function of channel widths $w_{c}$, $w_{s}$ and $w_{o}$ 
\cben{which correspond to the mean mixing times obtained when fixing the value of $w_{c}$, $w_{s}$ and $w_{o}$, respectively, and let the other widths vary. Width variations have a moderate effect on mixing times.}
\label{widr}} 
%\textbf{b)} Top view of the domain $\Omega_{s}$ and parameterization of the microfluidic mixer considered for the optimization process.}
\end{figure}

The value $MTE_{w_{c}}(i)$ corresponds to the mean values of the mixing times $J(\phi_{i,j,k})$ obtained when considering $j=1,..,4$ and $k=2,..,18$. The evolution of $MTE_{w_{c}}$, $MTE_{w_{c}}$ and $MTE_{w_{c}}$ are depicted in Figure \ref{widr}.

As we can observe in Table \ref{widt}, the most sensitive widths are the side inlets and central inlet with a mean mixing time variation of about 65\%. This result is expected, since those inlets carry  the denaturant solution and buffer flows, and thus affect the amount of injected products. Significant changes to these inlet geometries should be accompanied by changes in inlet velocities and performing a new optimization process as in Ref. \onlinecite{POF1}. For example, we hypothesize that variations which aim to preserve flow rate ratios should be explored first. On the other hand, outlet widths in the interval [2,13] $\mu$m (from the optimal value of 6 $\mu$m) will affect mixing time variation by only 7\%. Hence, we conclude that such errors on width have only a slight to moderate effect on mixing times.   Furthermore, regarding Figure \ref{widr}, we see that the mean mixing time is lower when considering values of $w_{c}$ and $w_{s}$ in the interval [2$\mu$m,4$\mu$m] and $w_{o}$ $\in$ [1$\mu$m,12$\mu$m]. Moreover, we remark that configurations with smaller inlets and bigger outlet are the worst from an efficiency point of view.

\begin{table}
\begin{center}
\caption{Maximum, minimum and mean values of the mixing time percent variation named $WE$, defined in Section \ref{widthe}, for the widths $w_c$, $w_s$ and $w_o$.} 
\label{widt}
\vspace{0.5cm}
\begin{tabular}{|c|ccc|}
\hline
\backslashbox{\textbf{Width}}{$\mathbf{WE}$}	&Mean	&Min	&Max\\
\hline
\textbf{$\mathbf{w_c}$}	&60	&27		&103	\\
\textbf{$\mathbf{w_s}$}	&68 	& 96	&118	\\
\textbf{$\mathbf{w_o}$}	&7	&1	&57\\
\hline
\end{tabular}
\end{center}
\end{table}

%--------------------------------------------------------------------------
\subsubsection{Mixer depth}
Next, we analyzed the effects of the mixer depth (in Z-direction).  Imperfections in micro-fabrication of these mixers can result in depth variations of approximately $\pm 1\mu$m \cite{refmixer}.  We thus computed the mixing time for mixers generated by considering the set of parameter $\phi_o$ and depths of $8,9,11$ and $12$ $\mu$m. The resulting mixing times (and their associated percent variation regarding the mixing time of the original mixer with a depth of 10 $\mu$m) were  0.14 $\mu$s (34\%), 0.12 $\mu$s (13\%),  0.10 $\mu$s (6\%), and 0.09 $\mu$s (13\%), respectively. 

As shown by these results, perturbations of $\pm 1$ $\mu$m generate reasonable percent variations in the weighted mean mixing time between 6\% and 13\%. As described previously, this indicates that errors in the mixer depth due to manufacturing processes do not strongly affect mixing performance for these relatively deep ($\sim$10 $\mu$m) mixers.  Note that the highest channel depth yields the lowest mixing time (0.09 $\mu$s for a depth equal to 12 $\mu$m versus 0.14 $\mu$s for the 8 $\mu$m depth). This result is expected, as the so-called wall effect (i.e., where the no-slip condition at the top wall results in low velocity values near the wall and near the corner where the X-Y plane meets the Y-Z plane) reduces the mixing performance near the mixer walls. Again, we see that the optimal mixer design (minimum mixing time) is influenced strongly by changes in manufacturing process (namely in achieving high aspect ratio features with deep reactive ion etching).  Mixer designs with relatively high channel-depth-to-feature width ratios yield optimal results. In our study, the minimum channel width (near $y$ = -1.5 $\mu$m) was 1.1 $\mu$m.  

\subsubsection{Shape Symmetry \label{ssym}}

The last geometrical aspect analyzed during this work is the impact of perturbations in the symmetry of the mixer according to the plane $x=0$ (including nonsymmetric injections velocities) on the mixer characteristics.

To this end, we considered the right half (versus quarter) of the geometry.  We then randomly generated 100 nonsymmetric mixers by considering perturbations of the parameters from 0.5\% up to 50\% of the left side (respecting to $x=0$) of the mixer shape and by keeping the right side of the mixer shape to its optimal value. These mixers were then classified according to the deviation observed between the streamlines starting from $(x=0,y=0,z=0)\mu$m of the symmetric and nonsymmetric mixers at the time when the non perturbed symmetric streamline reach $\Gamma_e$. According to this classification, we then computed the mean mixing time for each category and compared it to the optimized mixer mixing time by considering the percent variation formula (\ref{err}). Deviations in the intervals [0,0.3] $\mu$m,  [0.3,0.6] $\mu$m,  [0.6,0.9] $\mu$m, [0.9,1.2] $\mu$m, [1.2,1.5] $\mu$m and greater than 1.5 $\mu$m  generated mean mixing time percent variation of 14\%, 64\%, 114\%, 237\%, 328\% and 542\%, respectively. 

As we can observe from those data, for deviations  below  0.3 $\mu$m, which correspond to parameter perturbations lower than 10\% in the symmetry of shape and injection velocities, the order of the mixing time was conserved with a mixing time variation of 14\%. For greater deviations, the mixing time was dramatically increased from 64\% up to 500\%. Thus, we recommend normalized symmetry errors of less than 10\% be achieved to ensure a mixing time close to the optimal value.

\subsection{Flow injection velocities}
We studied the influence of injection velocities on  mixing time. The optimized injection velocities obtained in Ref \onlinecite{POF1} were  $u_s=$5.2 m s$^{-1}$ and $u_c=$0.2 m s$^{-1}$ (equivalent to a ratio $u_c/u_s$=0.0389). For this, we considered the optimized mixer and varied its side injection velocity $u_s=$ from 0.5 m/s to 9.5 m/s, with a step size of 0.5 m/s. Then, we chose $u_s$ in order to achieve $u_c/u_s$ ratios in the set $\{25,50,75,100,250\}$\% (considered as typical values) of the optimal ratio. This part of our study required a total of 95 evaluations of our model.

\begin{figure}[!ht]
\centering
\psfig{file=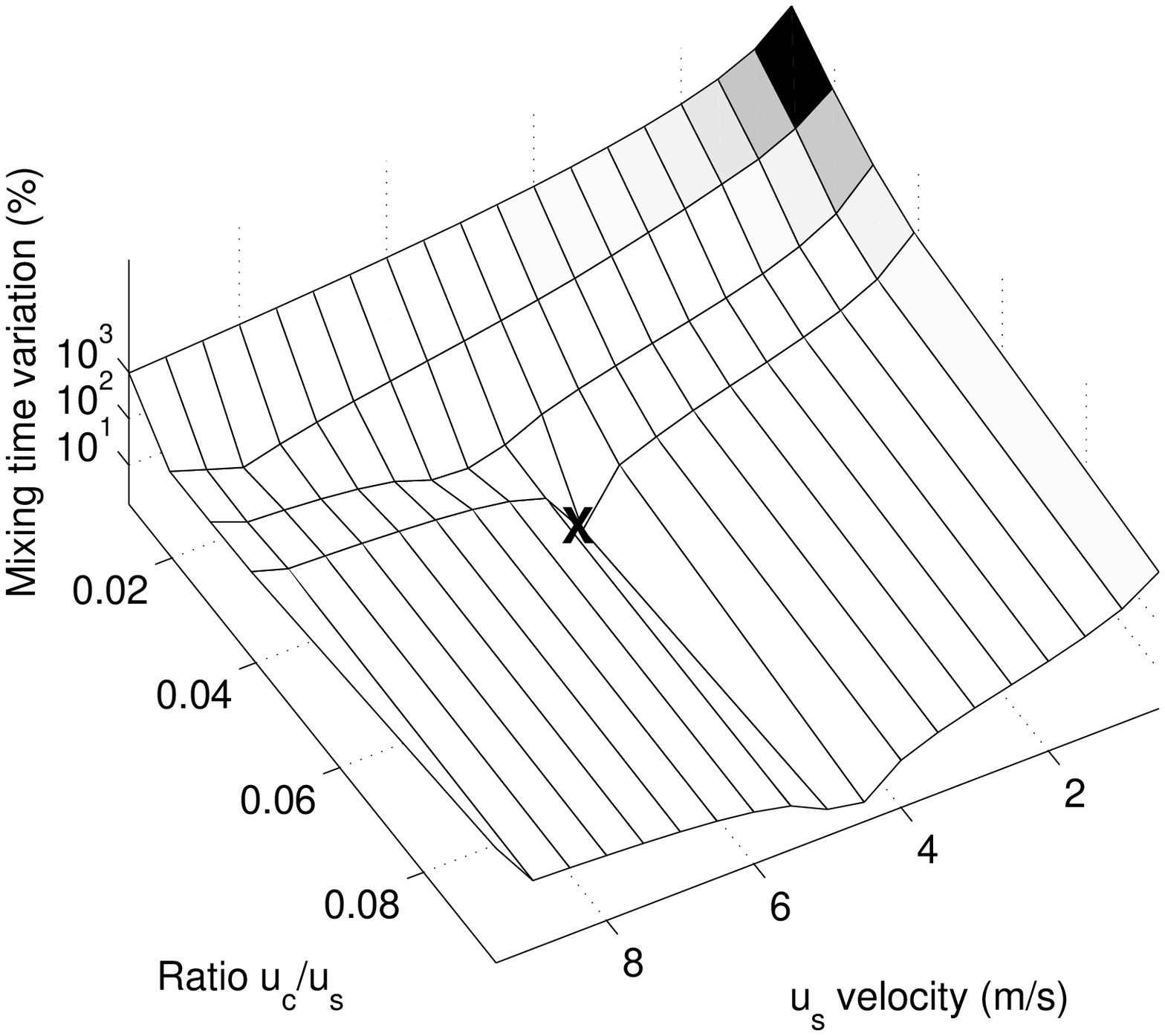,width=16cm} 
\caption{Percent variation of the mixing time relative to the optimal mixer (represented by X) as a function of variations of the side injection velocity $u_s$ (m/s) and the injection ratio $u_c/u_s$. Precise control of flow rates are crucial in mixing experiments. \label{velr}}  
%Variation in percent of the mixing time value relative to the optimal mixer (represented by X) and considering variations of the side injection velocity $u_s$ (m/s) and the injection ratio $u_c/u_s$.  \label{velr}} 
%\textbf{b)} Top view of the domain $\Omega_{s}$ and parameterization of the microfluidic mixer considered for the optimization process.}
\end{figure}

The results are summarized in Table \ref{velt} and Figure \ref{velr}. We can see that for $u_s$ $\in$ [4,9] m/s and $u_c$ $\in$ [0.12,0.88] m/s (i.e., a ratio $u_c/u_s$ of [0.0292,0.0973]), the mixing time has exhibited variations lower than 10\%. This suggests our mixer should be robust to small perturbations in the injection velocities.  In particular, the velocity of the central inlet flow should be in the interval [0.1,0.8]m/s to obtain a reasonable mixing time. Moreover, from those results we can deduce that if the ratio and/or $u_s$ are too small, the mixing time is drastically increased (more than 1000\%). In fact, the mixer performance becomes similar to the one achieved in a previous study (see Ref. \onlinecite{ijnme}).

\begin{table}
\begin{center}
\caption{Variation in percent of the mixing time value (relative to  the value of the optimal mixer indicated by -) as a function of values of the side injection velocity $u_s$ (m/s) and the injection ratio $u_c/u_s$.\label{velt}}
\vspace{0.5cm}
\begin{tabular}{|c||rrrrr|}
\hline
\backslashbox{\textbf{$\mathbf{u_c}$ (m/s)}}{\textbf{ratio}}  & 0.0097  &  0.0195 &   0.0292  &  0.0389 &   0.0973\\
\hline
\hline
0.5&26317&5444.8&1542.7&668.7&126.6\\
1&6136.1&1518.6&278.4&134.2&40.5\\
1.5&2776.7&667.4&109&56.8&20.8\\
2&1631.1&388.7&57.4&30.5&12.4\\
2.5&1083.7&229.5&34.6&18&7.8\\
3&775.2&142.6&22.2&11.1&4.9\\
3.5&583.3&91.1&14.8&6.6&3\\
4&456.8&59.7&9.8&3.6&1.5\\
4.5&368.9&41.6&6.2&1.8&0.4\\
5&304.6&30&3.6&-&0.5\\
5.5&256.6&22&1.6&1.4&1.3\\
6&219.5&16.4&1&2.4&1.9\\
6.5&190.4&12.2&1.1&3.2&2.4\\
7&166.8&9&2.1&3.9&2.9\\
7.5&147.5&6.4&2.9&4.5&3.3\\
8&131.6&4.4&3.6&5&3.6\\
8.5&118.1&2.8&4.2&5.4&3.9\\
9&106.7&5&4.7&5.7&4.2\\
9.5&97&9.2&9.4&10&42\\
\hline
\end{tabular}
\end{center}
\end{table}

We conclude that accurate control of flow rates is crucial to achieving fast mixing.  We recommend that flow rates be analyzed by experimental quantitation of inlet velocities using, for example, micron-resolution particle image velocimetry (as performed by Hertzog et al. in Ref. \onlinecite{refmixer}).

\subsection{Thermophysical Parameters}
We next studied the stability of the mixing time of the optimized mixer to changes in the thermophysical coefficients of the denaturant solution or in the concentration values needed to control the folding process. In physical experiments, these changes may result from uncertainties in conditions or solution properties (temperature, pressure, dilution, etc.)\cite{coef1} the following sections summarize.

\subsubsection{Denaturant Solution Parameters \label{denat}}
We chose for our work guanidine hydrochloride (GdCl) as a typical denaturant \cite{Dunbar,Kawahara} described by the following parameters: diffusivity in background buffer of  $D=2\times 10^{-9}$ m$^{2}$ s$^{-1}$, denaturant solution dynamic viscosity of  $\eta=9.8 \times 10^{-4}$  kg m$^{-1}$ s$^{-1}$ and mass density of $\rho=1010$ kg m$^{-3}$.

The thermophysical properties of GdCl solutions vary with concentration and ambient temperature:  consistent with the experimental work of Refs. \onlinecite{coef2}, \onlinecite{coef3} \cben{and \onlinecite{coef4}}, (i) the density of the GdCl \cben{solutions} can vary within  [1000,1700] kg m$^{-3}$; (ii) its viscosity can vary in [4,11]$\times 10^{-4}$ kg m$^{-1}$ s$^{-1}$; and (iii) its \cben{diffusivity} can vary in [1.9,13]$\times 10^{-9}$ m$^{2}$ s$^{-1}$. 

We considered the impact of these parameter variations on mixing time.  We varied each parameter within the aforementioned intervals using seven equispaced values. All possible configurations of parameters values were studied, which represents a total of 343 evaluations of the model. Then, similar to the work presented in Section \ref{widthe}, we computed the mean evolution of mixing time for each parameter value format both its lower and upper bound, and while varying the remaining coefficients to all their possible values.

\begin{figure}[htb]
\centering
\psfig{file=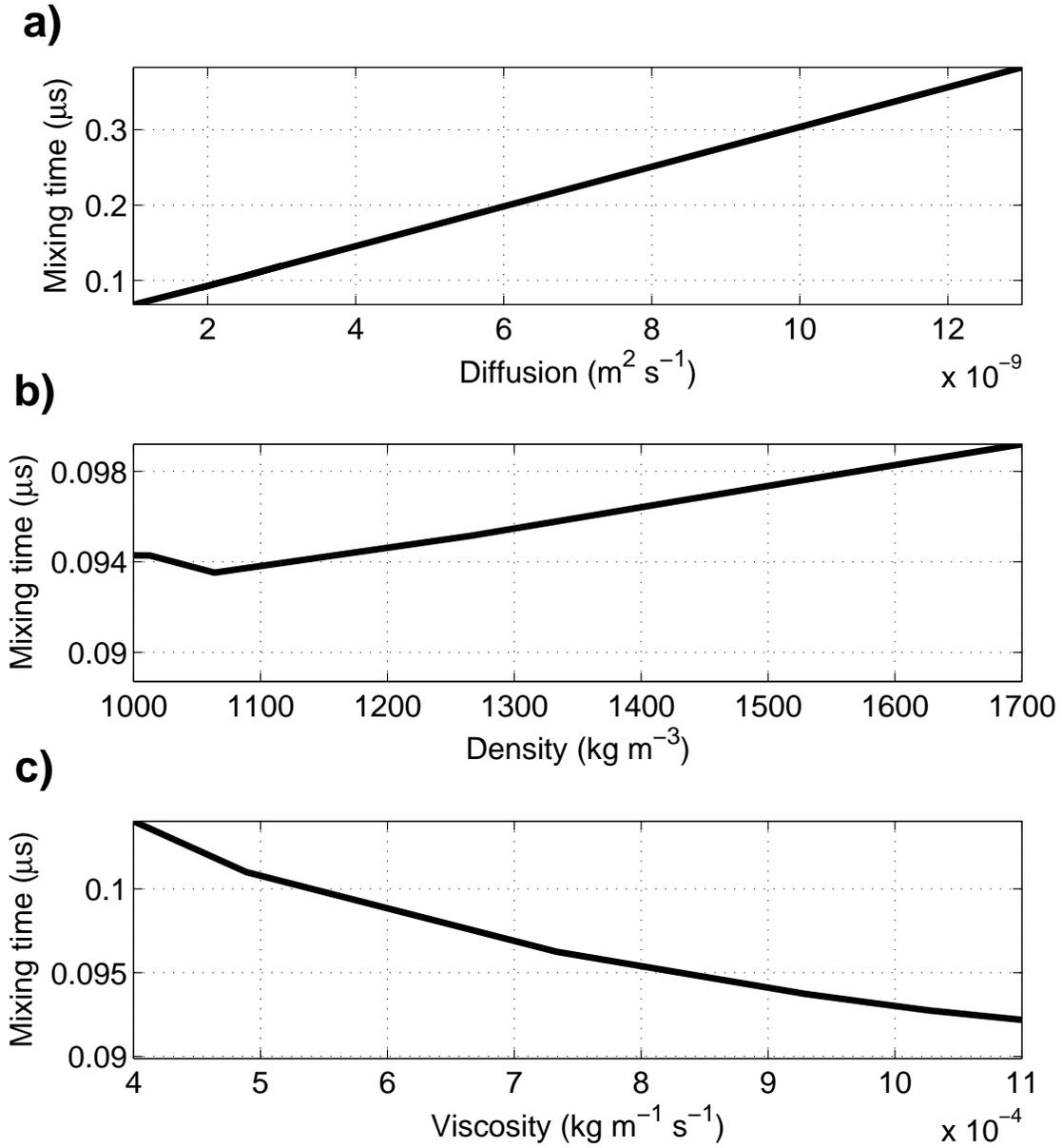,width=16cm} 
\caption{Mean mixing time (in s) \cben{obtained as a function of denaturant solution}  \textbf{a)} diffusivity, \textbf{b)} mass density and \textbf{c)} dynamic viscosity, and fixing the other two parameters. \label{coefr}  \label{cefrr}} 
%\textbf{b)} Top view of the domain $\Omega_{s}$ and parameterization of the microfluidic mixer considered for the optimization process.}
\end{figure}

The variations of the mean mixing time of the diffusion, density and viscosity are presented in Figure \ref{coefr}. We see that the diffusion was the most sensitive parameter, and can increase mixing time by up to 0.3$\mu$s. The other two coefficients maintained the mean mixing time close to 0.1$\mu$s. We note all of these values reasonable for the design as the order of mixing time is preserved.  We further note increasing viscosity and decreasing diffusivity and density result in lower mixing time.  The effect of decreasing diffusivity may at first seem counterintuitive, but mixing time is the result of a geometry- and flow-rate-dependent convective diffusion process.  For example, high diffusivity can result in significant decreases of denaturant concentration within the early-focusing region of the center jet, where fluid velocities are still too low to stretch material interfaces and decrease diffusion lengths of the center jet. The latter effect is discussed by Hertzog et al. (2004) (e.g., see Figure 2 of that reference).

\subsubsection{Concentration threshold \label{econc}}

Finally, we characterize the sensitivity of the mixer to the maximum and minimum denaturant concentration values of our mixing time (see (4)). The original mixer was designed to trigger unfolding for a concentration reduction of 60\%. We here consider mixing times for denaturant concentration reductions ranging from 10\% and  92\%. To this end, for a particular threshold value denoted by $\gamma$, we identify $\alpha_\gamma$ and $\omega_\gamma$ such that $\alpha_\gamma-\omega_\gamma=\gamma$ and they produce the minimum mixing time value  
\begin{equation}
J_{\gamma}(\phi)= \int_{c^{\phi}_{\omega_\gamma}}^{c^{\phi}_{\alpha_\gamma}} \dfrac{{\rm  d}y}{\mathbf{u}^{\phi}(y)},
\label{mtcf2}
\end{equation}
where   $c^{\phi}_{\alpha_\gamma}$ and $c^{\phi}_{\omega_\gamma}$ denote the Y-coordinates of the points situated along the streamline defined by the intersection of the two symmetry planes $z=0$ $\mu$m and $x=0$ $\mu$m, i.e. the y-axis, where the denaturant normalized concentration is  $\alpha_\gamma$ and $\omega_\gamma$, respectively.

\begin{figure}[htb]
\centering
\psfig{file=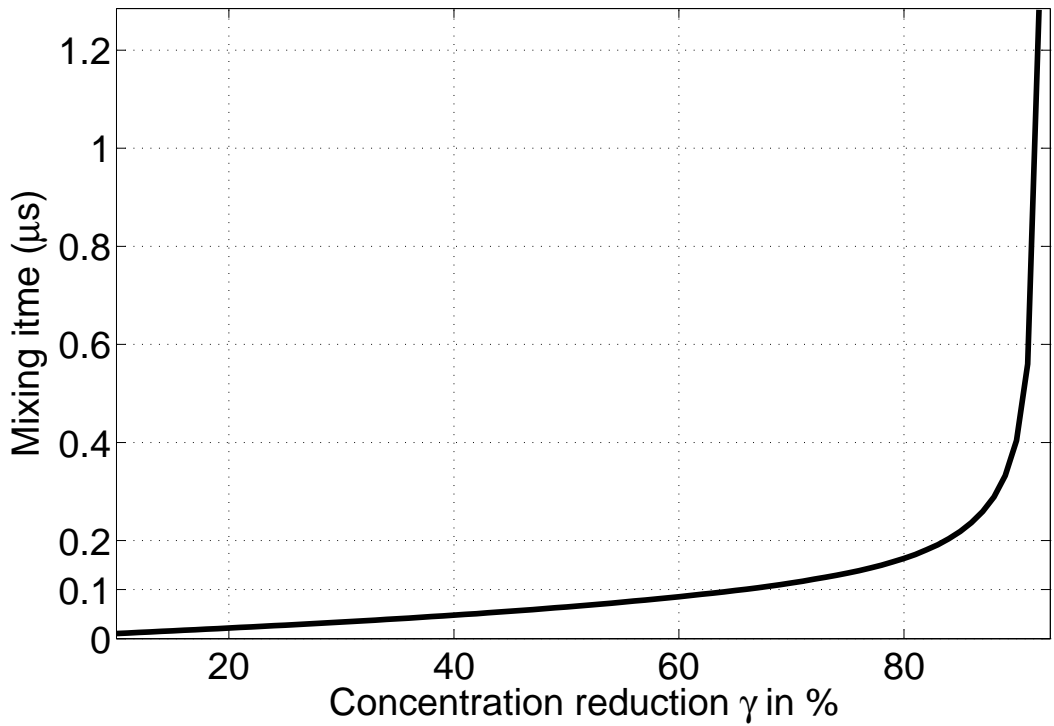,width=16cm} 
\caption{Mixing time (in s) as a function of the percentage reduction in concentration as  defined in Section \ref{econc}. \cben{The concentration values inherent to the definition of mixing have significant effect on mixing time. Note the} maximum concentration reduction allowed by complete mixing far downstream is $92\%$. \label{concr}} 
%\textbf{b)} Top view of the domain $\Omega_{s}$ and parameterization of the microfluidic mixer considered for the optimization process.}
\end{figure}

Results are presented in Figure \ref{concr}. The mixer exhibited mixing times lower than 0.4$\mu$s for up to a reduction of 90\%. We conclude that it is a robust design as the maximum reduction allowed by the flow rate ratios in this mixer was 92\%.  For a 70\% denaturant concentration reduction, we observed a 0.1$\mu$s mixing time.  The latter can be compared to the mixer of Ref. \onlinecite{POF1} which showed mixing times of 1$\mu$s for the same denaturant concentration reduction\cite{yao07}. 

\section{Conclusions}
We presented a detailed study of the robustness and performance of a microfluidic mixer design first presented in Ref. \onlinecite{POF1}.  The mixer is for protein folding dynamics studies and  can be used to initiate the folding process of a protein by diluting a local denaturant concentration in a short time interval. In Ref. \onlinecite{POF1}, we used a 3D numerical model and showed the ideal mixer shape, flow control parameters, and expected thermophysical parameters, which resulted in a mixing time of 0.10 $\mu$s.  We here studied the robustness of this mixer relative to expected variations of these major design features.   In particular, we studied (i) the uniformity of the mixing time through the center inlet flow and (ii) the sensitivity of the mixing time with respect to key mixer parameters. The uniformity study showed that mixing time is quite stable throughout the majority of the inlet stream (up to a distance from the walls of 0.4 $\mu$m).   With respect to design robustness, we found that the details of the mixer design in the region near the channel intersections are essential to the performance, i.e., the shape the minimum channel widths near this inlet,  the inlet flow velocity ratio, and possible (unwanted) asymmetries in the fabrication.  Other factors such as inlet channel angles, mixer depth (above a certain minimum), fluid properties, and denaturant concentration thresholds for protein folding have significantly weaker effect on mixing time.

Our analyses \cben{may} provide a guide to designers and fabricators of protein folding mixer devices, and can be used to evaluate trade-offs between manufacturing quality, precision of flow control, and expected performance.  Our work also serves as a case study associated with the general design and performance prediction of microfluidic devices, and may serve as a guide to designing complex and optimal fluidic systems.  In the least, the work highlights the complexity and importance of predicting and managing uncertainty in the performance of microfluidic systems.

In Table \ref{tabres}, for each parameter, we provide the mean, maximum and standard deviation values of the mixing time percentage variation regarding the optimal mixer obtained with this sensitivity analysis. %\textbf{NOTE: PLEASE COMPLETE THE ANALYSIS OF THIS TABLE IN ORDER TO GIVE RECOMMENDATIONS DURING THE MANUFACTURING PROCESS.}

\begin{table}
\begin{center}
\caption{Summary of  major findings of our  sensitivity analysis: Mean (\textbf{Mean}), Standard Deviation (\textbf{Dev}) and worst-case Maximum (\textbf{Max}) values of the mixing time in percentage of base design. For the sake of completeness, we also report the optimal value of each parameter (\textbf{Opt}) as well as the range of the considered values (\textbf{Range}). \label{tabres}} 
\vspace{0.5cm}
\begin{tabular}{|c|c|c|c|c|c|}
\hline
\textbf{Parameter}& \textbf{Opt} & \textbf{Range}	& \textbf{Mean} & \textbf{Dev} & \textbf{Max}\\
\hline
\multicolumn{6}{|c|}{\textbf{Intersection Angle}}\\
\hline
$\theta$	& $\pi$/5 & [0,2$\pi$/5]  &3	& 3 &16	\\
\hline
\multicolumn{6}{|c|}{\textbf{Channel Intersection}}\\
\hline
Prot$_{up}$	&0.8 	& [0,1]	&21& 17& 51	\\
Prot$_{low}$&0.7 &[0,1]&	30& 26 & 79\\
\hline
\multicolumn{6}{|c|}{\textbf{Channel Width}}\\
\hline
$\omega_c$ ($\mu$m)	&2 	& [1,4]	& 50 & 67 & 150	\\
$\omega_s$ ($\mu$m)	&3 	& [1,4]	& 53 & 57 & 181	\\
$\omega_o$ ($\mu$m)	&10 	& [2,18]	& 28 & 41 & 101	\\
\hline
\multicolumn{6}{|c|}{\textbf{Mixer Depth}}\\
\hline
Depth	 ($\mu$m) & 10 & [8,12]  &13 & 13&34	\\
\hline
\multicolumn{6}{|c|}{\textbf{Symmetry}}\\
\hline
Symmetry (\%) & 0 & [0.5,50]  &216	& 196&542	\\
\hline
\multicolumn{6}{|c|}{\textbf{Injection velocities}}\\
\hline
$u_c$ (m s$^{-1}$)	&0.2 	& [0.005,0.92]	& 68 & 133 & 304	\\
$u_s$ (m s$^{-1}$)	&5.2 	& [0.5,9.5]	& 51 & 153 & 669	\\
\hline
\multicolumn{6}{|c|}{\textbf{Physical coefficients}}\\
\hline
D (m$^2$ s$^{-1}$)	&2 $\times 10^{-9}$ 	& [1.9,13]$\times 10^{-9}$	& 155 & 198 & 307	\\
$\nu$ (kg m$^{1}$ s$^{-1}$)	&9.8 $\times 10^{-4}$ 	& [4,11]$\times 10^{-4}$	& 4 & 4 & 11	\\
$\rho$ (kg m$^{-3}$)	&1010 	& [1000,1700]	& 2 &2 & 6	\\
$\gamma$ (\%)	&60 	& [10,92]	& 129 & 271 & 1282	\\
\hline
\end{tabular}
\end{center}
\end{table}

\begin{acknowledgments}
This work was carried out thanks to the financial support of the Spanish “ Ministry of Economy and Competitiveness” under projects MTM2011-22658 and TIN2012-37483; the “Junta de Andaluc\'{\i}a” and the ”European Regional Development Fund (ERDF)” through projects P10-TIC-6002, P11-TIC-7176 and P12-TIC301; and the research group MOMAT (Ref. 910480) supported by "Banco Santander" and "Universidad Complutense
de Madrid". Juana López Redondo is a fellow of the Spanish ”Ram\'on y Cajal” contract program, co-financed by the European Social Fund.

\end{acknowledgments}

%\bibliographystyle{unsrt}
%\bibliography{microfluidic}% Produces the bibliography via BibTeX.

\end{document}